\documentclass{emulateapj}
\usepackage{apjfonts}
\usepackage{lscape}
\input{epsf}

\slugcomment{Accepted for publication in the Astrophysical Journal}

\shorttitle{Revised metallicity classes for low-mass stars}
\shortauthors{L\'epine, Rich, \& Shara}

\begin{document}

\title{Revised metallicity classes for low-mass stars: dwarfs
  (dM), subdwarfs (sdM), extreme subdwarfs (esdM), and ultra
  subdwarfs (usdM).\altaffilmark{1,2}}

\author{S\'ebastien L\'epine\altaffilmark{3}, R. Michael
  Rich\altaffilmark{4}, and Michael M. Shara\altaffilmark{3}}

\altaffiltext{1}{Based on observations conducted at the MDM
  observatory, operated jointly by the University of Michigan,
  Dartmouth College, the Ohio State University, Columbia University,
  and the University of Ohio.}

\altaffiltext{2}{Based on observations conducted at the Lick
  Observatory, operated by the University of California system.}

\altaffiltext{3}{Department of Astrophysics, Division of Physical Sciences,
American Museum of Natural History, Central Park West at 79th Street,
New York, NY 10024, USA, lepine@amnh.org, mshara@amnh.org}

\altaffiltext{4}{Department of Astrophysics, University of California
  at Los Angeles, Los Angeles, CA 90095, USA, rmr@astro.ucla.edu}

\begin{abstract}
The current classification system of M stars on the main sequence
distinguishes three metallicity classes (dwarfs - dM, subdwarfs - sdM,
and extreme subdwarfs - esdM). The spectroscopic definition of these
classes is based on the relative strength of prominent CaH and TiO
molecular absorption bands near $7000\AA$, as quantified by three
spectroscopic indices (CaH2, CaH3, and TiO5). The boundaries between
the metallicity classes were initially defined from a relatively small
sample of only 79 metal-poor stars (subdwarfs and extreme
subdwarfs). We re-examine this classification system in light of our
ongoing spectroscopic survey of stars with proper motion
$\mu>0.45\arcsec$ yr$^{-1}$, which has increased the census of
spectroscopically identified metal-poor M stars to over 400
objects. Kinematic separation of disk dwarfs and halo subdwarfs
suggest deficiencies in the current classification
system. Observations of common proper motion doubles indicates
that the current dM/sdM and sdM/esdM boundaries in the
[TiO5,CaH2+CaH3] index plane do not follow iso-metallicity contours,
leaving some binaries inappropriately classified as dM+sdM or
sdM+esdM. We propose a revision of the classification system based on
an empirical calibration of the TiO/CaH ratio for stars of near solar
metallicity. We introduce the parameter $\zeta_{TiO/CaH}$ which
quantifies the weakening of the TiO bandstrength due to metallicity
effect, with values ranging from $\zeta_{TiO/CaH}=1$ for stars of
near-solar metallicity to $\zeta_{TiO/CaH}\simeq0$ for the most
metal-poor (and TiO depleted) subdwarfs. We redefine the metallicity
classes based on the value of the parameter $\zeta_{TiO/CaH}$; and
refine the scheme by introducing an additional class of ultra
subdwarfs (usdM). We introduce sequences of sdM, esdM, and usdM stars
to be used as formal classification standards.
\end{abstract}



\section{Introduction}

Stars known as {\em red dwarfs} and {\em red subdwarfs}, are main
sequence stars of spectral type M, typically meant to include all
main sequence objects of spectral subtype K5 to M9. These cool,
low-mass stars are the most abundant stars in the Galaxy. They are
ubiquitous in the vicinity of the Sun, where they are readily
identified as faint stars with large proper motions. These stars also have
tremendous potential as tracers of the chemical and dynamical
evolution of the Galaxy, not only because of the sheer number of them,
but also because their lifetimes are much greater that the Hubble time
\citep{LBA97}, and because metallicity variations have dramatic
effects on their spectral energy distribution. The optical and near
infrared spectra of M dwarfs and subdwarfs are dominated by molecular
absorption bands of metal oxides and hydrides, most prominently bands
of TiO and CaH \citep{B91}. The ratio between the strength of the
oxide and hydride bands has long been known as a metallicity
diagnostic \citep{B82}. The high-velocity red subdwarfs, kinematically
associated with the Galactic halo, consistently display shallower TiO
absorption bands than the more common red dwarfs associated with the
Galactic disk. Atmospheric models show that metallicity variations in
cool and ultra-cool stars have dramatic effects on their optical
spectra \citep{AH95}. This is useful because low-resolution
spectroscopy is sufficient for measuring both effective temperature
and metallicity effects.

The current classification system of cool (spectral subtypes
K5-M6) and  ultra-cool (subtypes M7-M9) low-mass stars distinguishes
three broad metallicity classes: the dwarfs (K5-M9, or dK5-dM9), the
metal-poor subdwarfs  (sdK5-sdM9), and the very metal-poor extreme
subdwarfs (esdK5-esdM9). Dwarf (dM) stars are generally associated
with the Galactic disk population and show strong bands of both
TiO and CaH. Spectral subtypes from K5 to M9 are assigned based on a
sequence of standard objects \citep{KHM91}. The key feature is the
strength of the optical TiO bands, with VO bands also used to
discriminate the M7-M9 ``ultra-cool'' subtypes \citep{KHS95}. Spectral
subtypes of K and M dwarfs have also been calibrated relative to
spectral indices measuring the strengths of various molecular
bandheads. The most popular set of indices was defined by
\citet{RHG95}, and expanded by \citet{LRS03b}. These spectral indices
are useful in directly deriving spectral subtypes from simple
spectroscopic measurements.

The metal-poor subdwarfs (sdM) and extreme subdwarfs (esdM) are
for the most part kinematically associated with the Galactic halo, and
display relatively weaker bands of TiO while retaining strong CaH
absorption \citep{HCM84,RA93,GSIJ97,GR97a,LRS03b,SLMZ04,PB06}. A
classification system for subdwarfs was introduced by Gizis (1997,
hereafter G97), and is based on measurements of the four spectroscopic
indices (CaH1, CaH2, CaH3, and TiO5) defined by \citet{RHG95}. These
indices measure the strength of the main CaH and TiO bands located
near 7,000\AA. Separation into the three metallicity classes is
determined by the relative strengths of the CaH1, CaH2, and CaH3
spectral indices relative to the TiO5 spectral index, using
arbitrarily defined relationships. Subtypes are assigned based on the
combined strengths of the CaH2 and CaH3 indices. The system was
initially defined for earlier subtypes, down to $\sim$sdM5/esdM5, but
has since been extrapolated to cooler subtypes
\citep{SSSIM99,LSR03a,LSR04,SLM04}, and subdwarfs are now being
identified well into the L type regime \citep{LRS03c,Betal03,GH06}.

In the G97 system the separation between dwarfs, subdwarfs, and
extreme subdwarfs is determined using a combinations of
four relationships defined in [CaH1,TiO5], [CaH2,TiO5], and
[CaH3,TiO5]. Following \citet{LSR03a}, it has however become customary
to represent the spectroscopically confirmed subdwarfs in a single
diagram of CaH2+CaH3 versus TiO5 \citep{LSR04,SLMZ04,RG05}. A
simplification of the G97 system was thus proposed by \citet{BK06},
using only two relationships defining dM/sdM/esdM separators in the
[CaH2+CaH3,TiO5] diagram. The separators were validated using
previously classified objects, and are thus essentially equivalent to
the original G97 system.

In this paper, we re-examine the classification system for red dwarfs
and subdwarfs in light of new spectroscopic measurements of 1,983 high
proper motion stars from the LSPM-north proper motion catalog
\citep{LS05}. We redefine the metallicity subclasses based on a
calibration of the TiO to CaH ratio for stars of Solar metallicity. We
introduce the $\zeta_{\rm TiO/CaH}$ metallicity index, which we
demonstrate to be an simple and objective measurement of the metal
content in M dwarfs and subdwarfs. The index therefore provides an
objective criterion to redefine the metallicity subclasses. We propose
an update of the classification system which uses four metallicity
subclasses instead of the previous three, introducing the new subclass
of ``ultra subdwarfs'' (usdK, usdM) which includes stars with the
lowest apparent metal content. We introduce new spectroscopic
sequences of subdwarfs, extreme subdwarfs, and ultra subdwarfs, which
can be used as classification standards.

\section{Current census of spectroscopically confirmed M subdwarfs}

\begin{figure*}
\epsscale{0.9}
\plotone{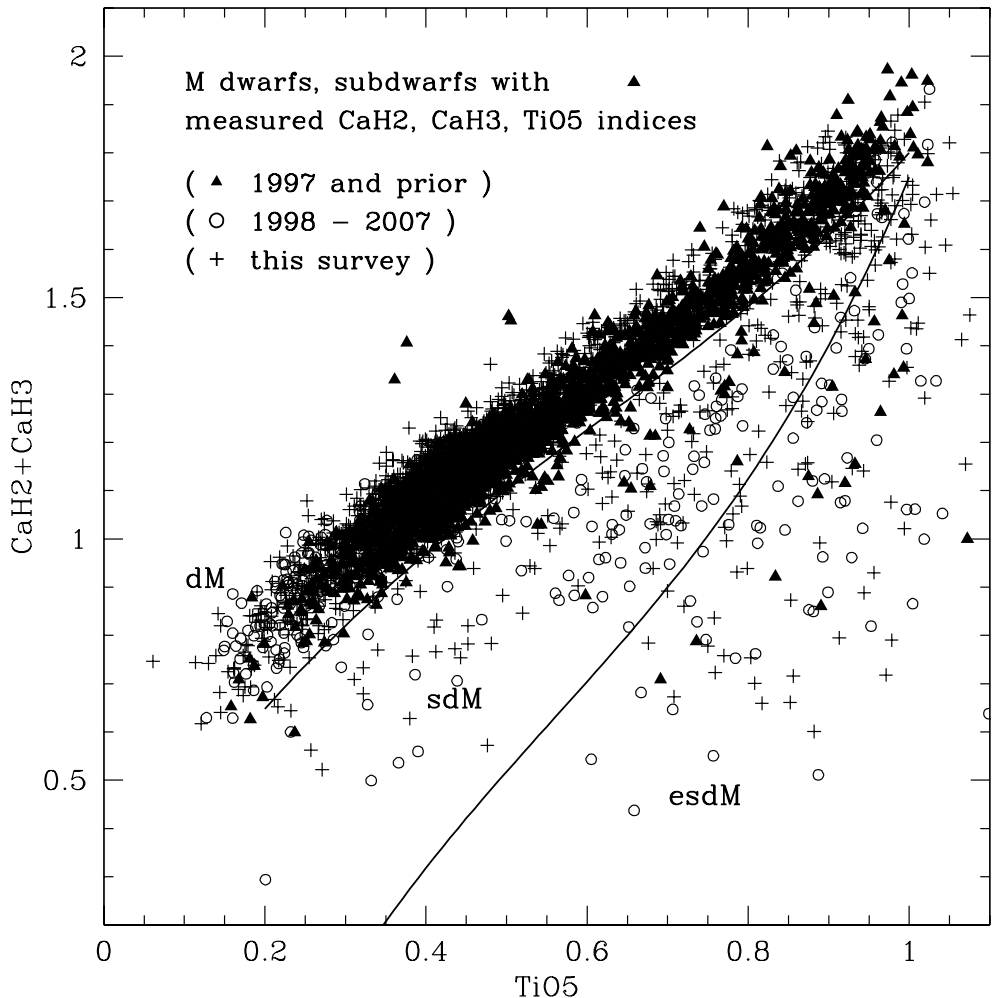}
\caption{Top: distribution of spectroscopically confirmed, cool dwarfs
  and subdwarfs in the [CaH2+CaH3,TiO5] spectral indices
  diagram. Stars with spectral indices measured up until 1997 are
  marked as filled triangles. Stars identified from 1998 to 2007 are
  marked as open circles. Objects classified in our current,
  spectroscopic follow-up survey of high proper motion stars from the
  LSPM-north catalog are denoted by crosses. The two continuous lines
  marks the dM/sdM and sdM/esdM subclass separators, as proposed by
  \citet{BK06}, which delimit the metallicity subclasses according
  to the system introduced by \citet{G97}.}
\end{figure*}

\subsection{Previously classified objects}

\begin{deluxetable}{lcc}
\tabletypesize{\scriptsize}
\tablecolumns{3} 
\tablewidth{0pt} 
\tablecaption{Spectral Indices for Low-Mass Subdwarfs\tablenotemark{a}}
\tablehead{
\colhead{Index Name} & 
\colhead{Numerator(\AA)} &
\colhead{Denominator(\AA)}
}
\startdata 
CaH2   & 6814-6846 & 7042-7046\\
CaH3   & 6960-6990 & 7042-7046\\
TiO5   & 7126-7135 & 7042-7046 
\enddata
\tablenotetext{a}{Indices originally defined by \citet{RHG95}; the
  indices are calculated from the ratio of the {\em average flux} over
  the specified wavelength ranges.}
\end{deluxetable} 

The system currently in use to classify the M subdwarfs is based on
the strength of the TiO and CaH molecular bands near 7000\AA, as
quantified by the three spectral indices CaH2, CaH3, and TiO5, defined
by \citet{RHG95}. The CaH2 and CaH3 indices measure the mean flux
level in parts of the main CaH band, as measured relative to the mean
flux in a reference point set at 7044\AA (which defines a
local pseudo continuum). The TiO5 index measures the mean flux level
in a region within the nearby TiO band, as measured relative to the
same pseudo-continuum reference point. The formal index definitions
are recopied in Table 1. The location of all the reference points is
illustrated in Figure 1 of \citet{RHG95} and in Figure 1 of
\citet{LSR03a}. The indices are defined in such a way that they yield
a value of 1 if the band is undetected (i.e. the depth of the bandhead
is at the continuum level), and 0 if the band is completely saturated.

We have searched the literature for spectroscopically confirmed M
dwarfs and subdwarfs, for which values of the CaH2, CaH3 and TiO5
indices have been measured and recorded. These include most of the
stars classified as M dwarfs in the Solar Neighborhood ($d<25$pc), and
all the spectroscopically confirmed M subdwarfs. The distribution in
the [CaH2+CaH3,TiO5] diagram is shown in Figure 1 (filled triangles,
open circles).

Historically the first set of objects for which values of CaH2, CaH3
and TiO5 were recorded are the 1,971 nearby dwarfs from the
Palomar-MSU spectroscopic survey \citep{RHG95}. High proper motion
stars from W. J. Luyten's {\em LHS catalog} were then targeted: 71 K
and M dwarfs were classified by \citet{GR97a}, then 79 K and M
subdwarfs were classified by \citet{G97}. These stars were the
fundamental sample from which \citet{G97} established the M subdwarf
classification system based on the molecular band indices. Stars from
this initial sample are noted with filled triangle symbols in
Fig.1. The census of spectroscopically confirmed M subdwarfs and
extreme M subdwarfs was then relatively small, and many additions have
been made since then.

Most of the new additions came from follow-up spectroscopy of faint
stars with large proper motions, including objects listed in the {\em
 LHS catalog}, and newly discovered stars from the SuperCosmos sky
survey \citep{HDIM01}, the Liverpool-Edinburgh proper motion survey
\citep{PJH03}, and the SUPERBLINK proper motion survey \citep{LS05}.
Some of the newest additions have been pushing the limits of the
classification system to ever cooler subdwarfs. The latest objects in
the G97 system were LHS 377, an sdM7.0 subdwarf, and LHS 1742a, an
esdM5.5 extreme subdwarf; much later stars have been identified
since. The high proper motion star APMPM J0559-2903 was classified an
esdM7.0 by \citet{SSSIM99}, while LSR 1425+7102 was classified an
sdM8.0 \citep{LSR03a}. The very high proper motion object LSR 
1610-0040 was initially classified as an sdL by \citet{LRS03c}, though
the star is now considered to be a moderately metal-poor late-M
dwarf/subdwarf \citep{CV06}. \citet{SLMZ04} identified three more very
cool subdwarfs, including SSSPM J1013-1356, classified as
sdM9.5. \citet{Lod05} classified the new high proper motion star SSSPM
J0500-5406 an esdM6.5, though values of the CaH2, CaH3 and TiO5
indices were not included in the paper. They were, however, provided
by \citet{BK06}, who re-observed SSSPM J0500-5406 and identified two
more ultra-cool subdwarfs including LEHPM 2-59, classified as
esdM8.0. The 3.5$\arcsec$ yr$^{-1}$ proper motion star SSSPM
J1444-2019 was classified as an sdM9 \citep{SLM04}. In all these
cases, the spectral subtypes were determined by extrapolating the
relationships defined by G97, in which the spectral subtype is a
monotonic function of the CaH2 and CaH3 indices. Other late-type
subdwarfs reported in the literature include the high proper motion
star APMPM J1523-0245, classified as sdM5.5 \citep{GSIJ97}, and a few
late-M and L subdwarfs identified within the 2MASS survey
\citep{Betal03,Betal04a,B04b}.
 
More extensive lists of new classifications included 257 dwarfs and
subdwarfs from the {\em LHS catalog} \citet{RG05}, and  104 M/sdM/esdM
stars from the SUPERBLINK survey \citep{LRS03b}. All additions to the
census of spectroscopically confirmed M dwarfs and subdwarfs made
between 1998 and 2007 are displayed as open circles in Fig.1. The
figure  also shows the new dM/sdM and sdM/esdM separators as defined
by \citet{BK06}. The diagram demonstrates the relative scarcity of
subdwarfs and extreme subdwarfs in the Solar Neighborhood, which are
outnumbered by the dwarfs by two orders of magnitude.

\subsection{New spectroscopic survey}

We have been conducting a massive spectroscopic follow-up survey of
sources from the LSPM-north catalog of stars with proper motions
$\mu>0.15\arcsec$ yr$^{-1}$ \citep{LS05}. About half the stars in the
LSPM-north catalog are re-identifications of stars from the Luyten
catalogs, the other half are new discoveries, and the LSPM-north is
thus an excellent source of new nearby M dwarfs and
subdwarfs. Medium-resolution spectra with
$\Delta\lambda\sim1.7-2.3\AA$ pixel$^{-1}$ have been collected from
the 2.4-meter Hiltner telescope at the MDM Observatory, the 3-meter
Shane Telescope at the Lick Observatory, and the 4-meter Mayall
Telescope on Kitt Peak. Our current target list includes all 2,966
stars from the LSPM-north which have proper motions $\mu>0.45\arcsec$
yr$^{-1}$, with a priority to stars with no published spectroscopic
data. Detailed results from the ongoing survey will be published in an
upcoming spectroscopic catalog (L\'epine et al., {\it in
  preparation}). To this date, we have identified 1,983 stars in our
survey which clearly display the broad, molecular absorption bands
characteristic of low-mass, main sequence stars with spectral subtypes
from K5 to M9, including significant numbers of subdwarfs and
extreme subdwarfs.

We have been recording values of the CaH2, CaH3, and TiO5 indices for
all the stars. With spectra of resolution 1.5\AA-2.3\AA per pixel, and
signal-to-noise S/N$\sim20$, our index measurements have a typical
accuracy of $\approx0.03$. A significant fraction of the stars,
especially the subdwarfs, are found to have large radial velocities
(consistent with Galactic halo membership); all spectra are thus red-
or blue-shifted to their local rest frame before the spectroscopic
indices are measured.

Our sample of newly classified dwarfs and subdwarfs spans a wide
variety of relative and absolute CaH and TiO bandstrengths, which
indicates that the sample covers a wide range of both effective
temperatures and metal abundances. The newly classified stars are
plotted in Fig.1 as crosses. The new sample fills up many holes left
in the [CaH2+CaH3,TiO5] diagram by the previous census. Our follow-up
survey thus represents a major addition to the list of
spectroscopically confirmed M subdwarfs, and provides a much expanded
sample of stars with recorded values of the CaH2, CaH3 and TiO5
indices.

The number of spectroscopically confirmed subdwarfs keeps increasing
at a rapid pace. We expect current and planned surveys to
significantly increase the number of spectroscopically confirmed
subdwarfs. With a large dataset in hand, a re-evaluation of the
classification system is now in order.

\section{Revisiting the metallicity subclasses of low-mass stars}

\subsection{Dwarf/subdwarf separation in the reduced proper motion
  diagram}

It has long been known that the high-velocity subdwarfs in the Solar
vicinity, i.e. local stars from Population II, can be separated out
from Galactic disk (Population I) stars using a reduced proper motion
diagram \citep{J71}. A plot of the reduced proper motion against an
optical-to-infrared color is known to provide a relatively clean
separation between M dwarfs from the local disk population and M
subdwarfs from the Galactic halo \citep{SG02}. It is thus possible to
isolate samples of Pop I dwarfs and Pop II subdwarf using reduced
proper motion diagrams. Those samples can then be used to determine the
expected distribution of metal-rich and metal-poor stars in the
[CaH2+CaH3,TiO5] diagram, and objectively determine the boundary
separating the M dwarfs and M subdwarfs.

\begin{figure*}
\epsscale{1.0}
\plotone{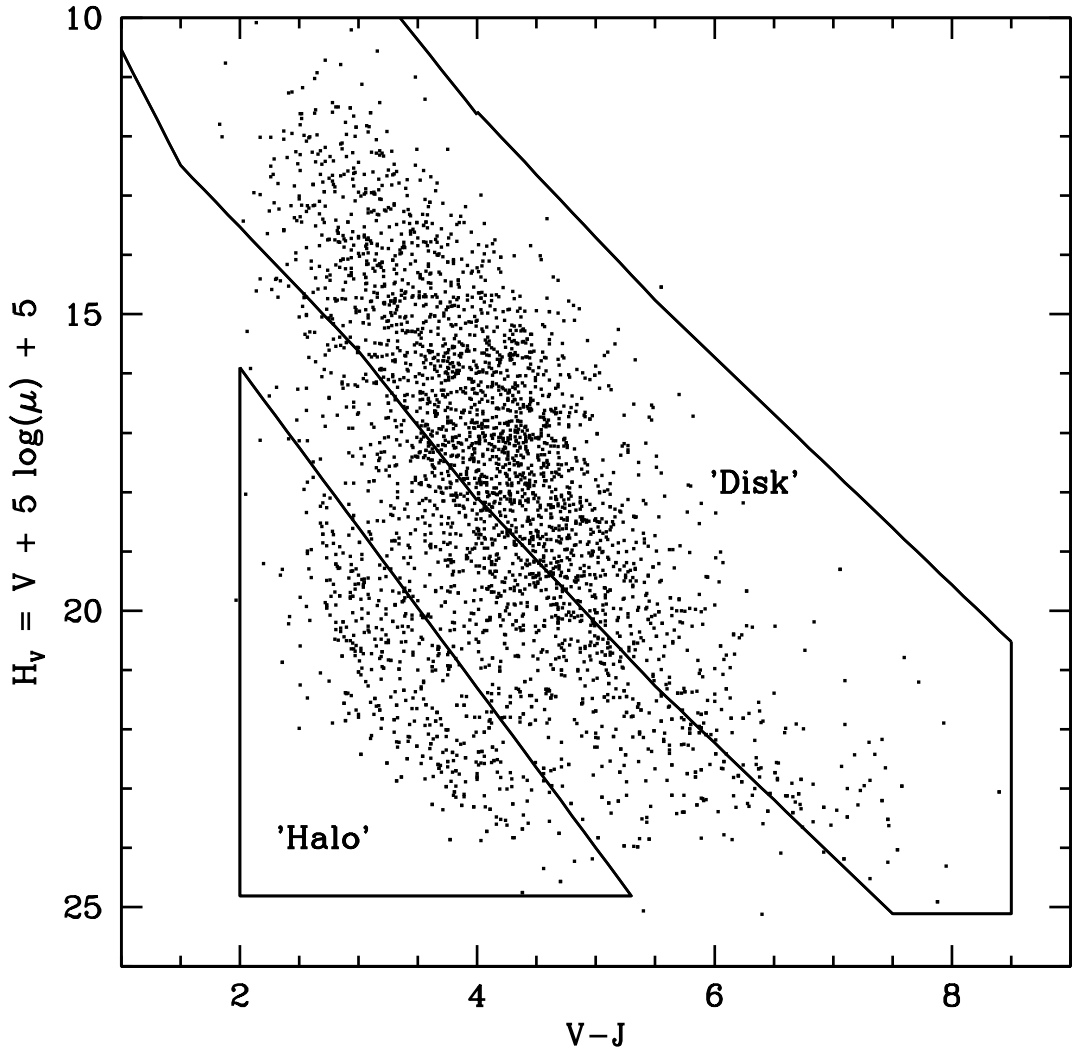}
\caption{Reduced proper motion diagram of the M dwarfs and subdwarfs
  with formal spectral classification (see Fig.1). We identify two
  distinct groups: the ``disk'' stars, with generally low transverse
  velocities and standard color-magnitude relationship, and the
  ``halo'' stars with generally large transverse velocities and
  fainter magnitudes at a given color.}
\end{figure*}

We build a reduced proper motion diagram using the reduced proper
motion in V magnitude $H_V=V+5\log{\mu}+5$ and the optical-to-infrared
color V-J. The diagram is shown in Figure 2, and displays all the
dwarfs and subdwarfs for which we could recover proper motions
($\mu$), V magnitudes, and V-J colors. As expected, the stars fall in
two fairly distinct loci, which correspond to the metal-rich disk and
metal-poor halo stars, respectively. As is well-known, halo subdwarfs
are separated from the disk M dwarfs both because of their relatively
bluer color at a given absolute magnitude \citep{Metal92}, and because
of their larger mean transverse velocities. Both effects combine to
move the M subdwarfs down and to the left of the main stellar locus.

We define two regions, denoted "Disk" and "Halo" which, based on the
argument above, should contain a majority of metal-rich/disk stars and
metal-poor/halo stars, respectively. One can show that the reduced
proper motion $H_V = M_V + 5 \log{v_T} - 3.38$, where
$M_V$ is the absolute magnitude in $V$ and $v_T$ is the projected
velocity in the plane of the sky, in km s$^{-1}$. We assume the
absolute magnitudes M$_V$ to follow the color-magnitude relationship
M$_V$=M$_V$($V-J$) of \citet{L05}, which was calibrated using a large
sample of nearby dwarfs with accurate parallax measurements.
The ``Disk'' region is hence defined in the [$H_V$,$V-J$] plane to
includes all dwarfs with transverse velocities 5 km s$^{-1} < v_T < $
100 km s$^{-1}$. Stars which fall below the ``Disk'' box in Fig.2 are
thus either dwarfs with transverse velocities $v_T >$ 100 km s$^{-1}$,
or stars which fall significantly below the M$_V$=M$_V$($V-J$)
color-magnitude relationship for nearby disk dwarfs, e.g. subdwarf
stars. The ``Halo'' region, on the other hand, selects for stars which
are unambiguously high-velocity, Pop II stars. The upper boundary
in Fig.2 selects for cool subdwarfs with transverse velocities
$\gtrsim150$ km s$^{-1}$. The many stars which fall in the no man's
land between the ``Disk'' and ``Halo'' regions probably include
moderately metal-poor subdwarfs, or metal-rich stars with large
transverse velocities. Such stars are likely to be associated with the
Galactic thick disk.

\begin{figure*}
\epsscale{1.0}
\plotone{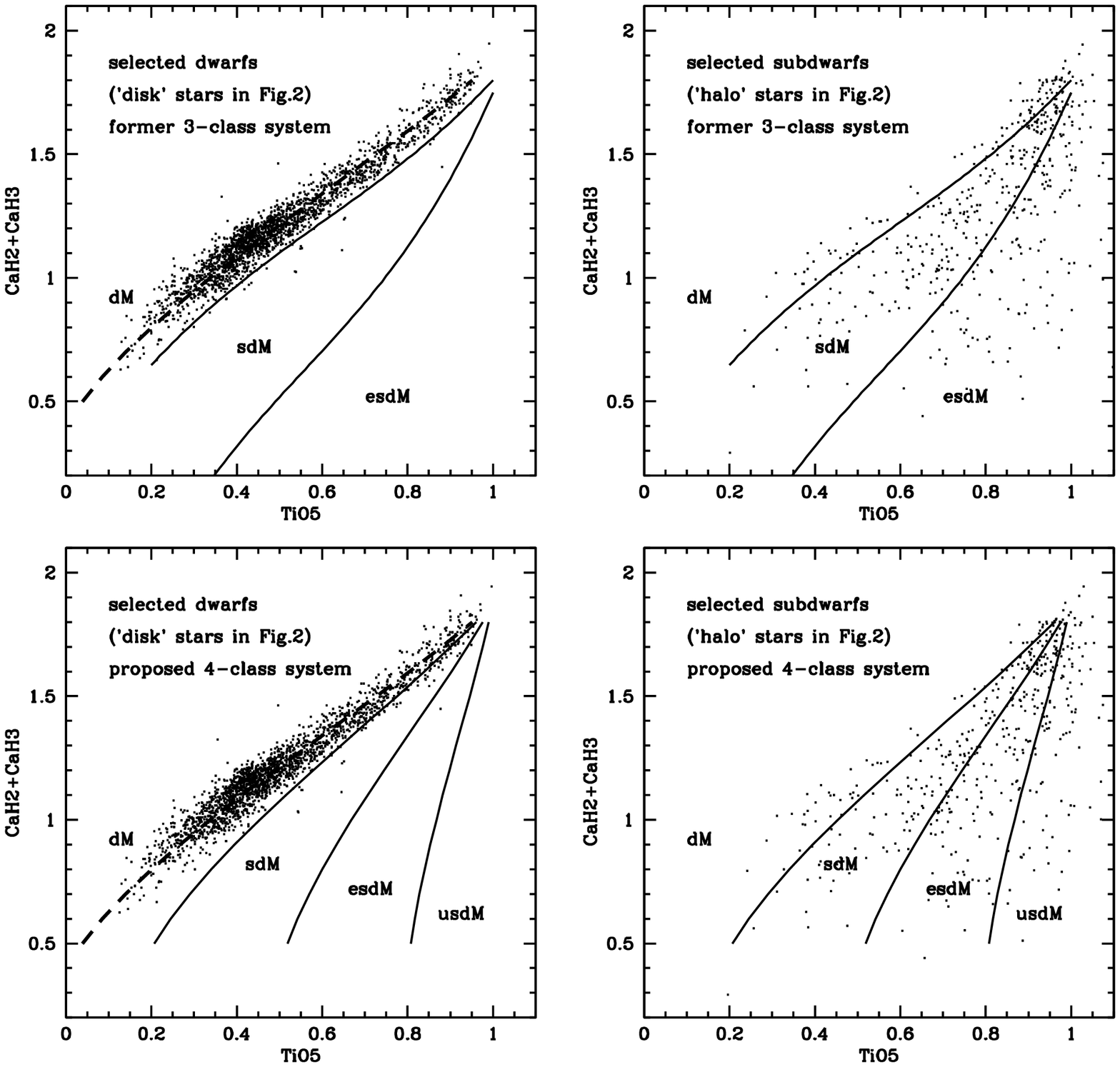}
\caption{Top left: Comparison between the CaH and TiO bandstrengths
  of the stars in the ``Disk'' group, as defined in Fig.2. The TiO and
  CaH bandstrengths are tightly correlated, which indicates that the
  stars have equivalent (Solar) metallicities. Top right: CaH and
  TiO bandstrength for stars in the ``Halo'' group (see Fig.2). The
  TiO bands are systematically weaker than those of the ``Disk''
  groups, which confirms that they are indeed drawn from a metal-poor
  population (Pop II). Stars are also found with a broad range of
  TiO/CaH ratios, which indicates that they span a wide range of
  metal abundances. Bottom left: new separators for the proposed
  4-class system, which are based on the calibration of the
  CaH2+CaH3/TiO5 relationship for disk dwarfs (dashed line). Bottom
  right: distribution of the ``Halo'' stars in the 4-class system.}
\end{figure*}

Figure 3 shows separate [CaH2+CaH3,TiO5] diagrams for stars which fall
into either the ``Disk'' or ``Halo'' selection boxes. Stars from the
``Disk'' group show a tight correlation between the CaH2+CaH3 and TiO5
indices (Fig.3, left panels). In contrast, stars from the "Halo" group
display a wide variety of TiO bandstrengths for a given depth of the
CaH bands. The ``Halo'' stars also have systematically smaller ratios
of TiO to CaH, compared with the stars from the ``Disk'' group. This
confirms our earlier impression that the kinematically selected
``Disk'' objects are indeed dwarf stars all of roughly solar
metallicity, while the ``Halo'' objects are overwhelmingly metal-poor
subdwarfs.

Assuming this scheme to be correct, then the two groups can be
used as fiducial samples to more precisely determine the loci of dwarfs
and subdwarfs in the [CaH2+CaH3,TiO5] diagram. This is especially
useful to redefine the separator between the dwarf (M) and subdwarf
(sdM) classes. A close examination of the top panels in Fig.3 shows
that the current M/sdM separator is fairly conservative in assigning
stars to the subdwarf (sdM) class. The vast majority of the ``disk''
stars (top left panel) are allocated to the dwarf (dM) class, with
room to spare. Many stars from the ``Halo'' group, on the other hand,
are not assigned to the subdwarf (sdM) class, as they most probably
should be. This is especially apparent at earlier subtypes,
i.e. for larger values of CaH2+CaH3 (top right of the diagram). This
justifies at least a readjustment of the metallicity class separators
proposed by \citet{BK06}.

\subsection{The metallicity index $\zeta_{\rm TiO/CaH}$}

The kinematical selection of local disk dwarfs from the reduced proper
motion diagram allows us to calibrate the ratio of TiO to CaH for
stars of the disk, which contains stars of roughly solar
metallicity. We use all the stars from the ``disk'' group, and obtain
a fit of the TiO5 spectral index as a function of the CaH2+CaH3 index,
yielding:
\begin{displaymath}
[{\rm TiO5}]_{Z_{\odot}} = -0.164 ({\rm CaH2+CaH3})^3 + 0.670 ({\rm
  CaH2+CaH3})^2 
\end{displaymath}
\begin{equation}
- 0.118 ({\rm CaH2+CaH3}) - 0.050 
\end{equation}
Equation 1 effectively provides a calibration of the TiO
bandstrength relative to the CaH bandstrength, for stars of roughly
solar metallicity ($Z_{\odot}$). This defines an objective reference
point from which to estimate atmospheric abundances in subdwarfs spectra.
While the depth of the CaH bands is used to estimate the effective
temperature, the ratio of TiO to CaH is used to estimate the
metallicity. A curve denoting $[{\rm TiO5}]_{Z_{\odot}}$ as a function
of CaH2+CaH3 is drawn in Fig.3 (dashed line). The ``disk'' stars are
found tightly clustered along this curve, while the ``halo'' stars
span a large range of TiO5 values with $[{\rm
    TiO5}]_{Z_{\odot}}\lesssim$TiO5$\lesssim1.0$.

The relative strength of TiO to CaH in any one star can then be
expressed relative to the (calibrated) strength of the TiO band in
stars of solar metallicity using the parameter $\zeta_{\rm TiO/CaH}$,
which we define as:
\begin{equation}
\zeta_{\rm TiO/CaH} = \frac{1 - {\rm TiO5}}{1 - [{\rm TiO5}]_{Z_{\odot}}},
\end{equation}
where $[{\rm TiO5}]_{Z_{\odot}}$ is given by Eq.1, and is a function of
the CaH2+CaH3 index. The parameter $\zeta_{\rm TiO/CaH}$ can thus be
calculated for any star from its CaH2, CaH3, and TiO5 indices. All
stars with abundances equal to the average abundance of stars in the
Galactic disk will have $\zeta_{\rm TiO/CaH}=1$, by
definition. Metal-poor stars will have $\zeta_{\rm TiO/CaH}<1$, while
stars with metal abundances larger than the Galactic disk average will
have $\zeta_{\rm TiO/CaH}>1$. 

\begin{figure}
\epsscale{3.0}
\plotone{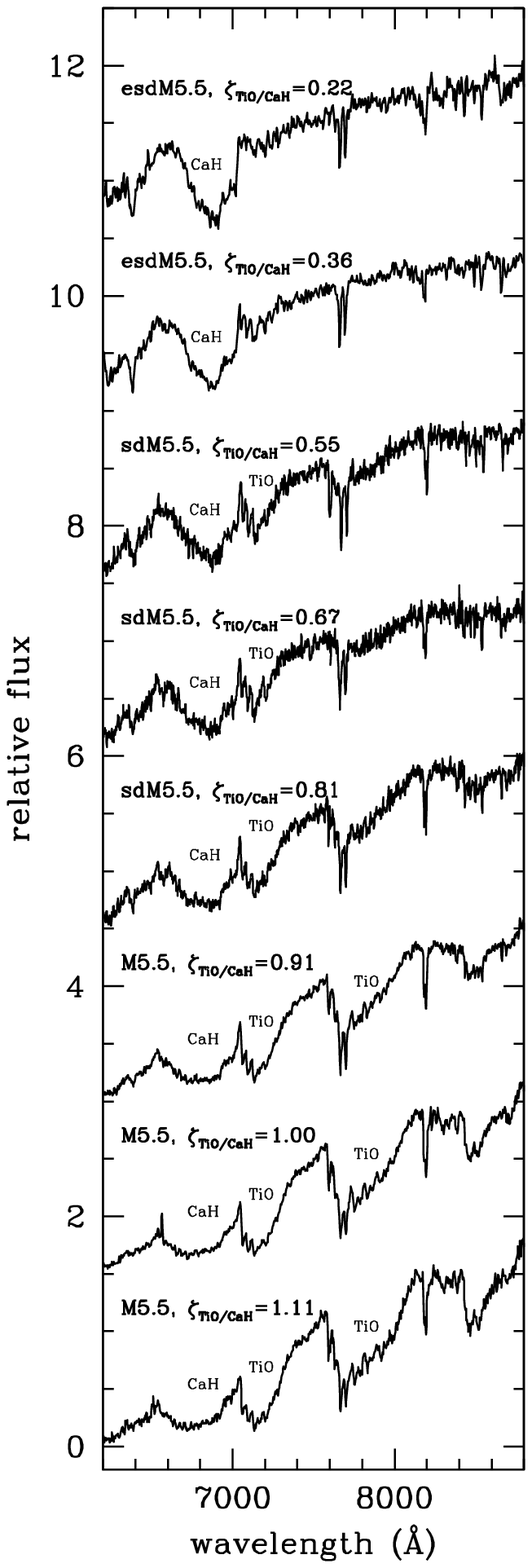}
\caption{Effects of metallicity on the spectra of cool
  dwarfs and subdwarfs. These low-mass stars of similar [CaH2+CaH3]
  index values are shown in order of increasing $\zeta_{\rm TiO/CaH}$,
  from top to bottom. The metal abundance is reflected in the relative
  strength of the TiO bandhead to the CaH bandhead. Spectral subtypes
  are noted, as well as values of the $\zeta_{\rm TiO/CaH}$ index which
  quantifies the ratio between the TiO and CaH bandstrengths and is
  thus a marker for metallicity effects.}
\end{figure}

Figure 4 displays a sequence of objects with similar CaH
bandstrengths, but spanning a wide range of TiO bandstrengths, as
determined by the CaH2+CaH3 and TiO5 spectral indices. Values of
$\zeta_{\rm TiO/CaH}$ are noted for each star, along with spectral
subtypes. The stars clearly follow the expected trends, with the
extreme subdwarfs (esdM) having the smallest values of $\zeta_{\rm
  TiO/CaH}$, and the M dwarfs the largest. The spectra in Fig. 4
therefore trace metallicity effects in M dwarfs and subdwarfs.

\begin{figure}
\epsscale{1.1}
\plotone{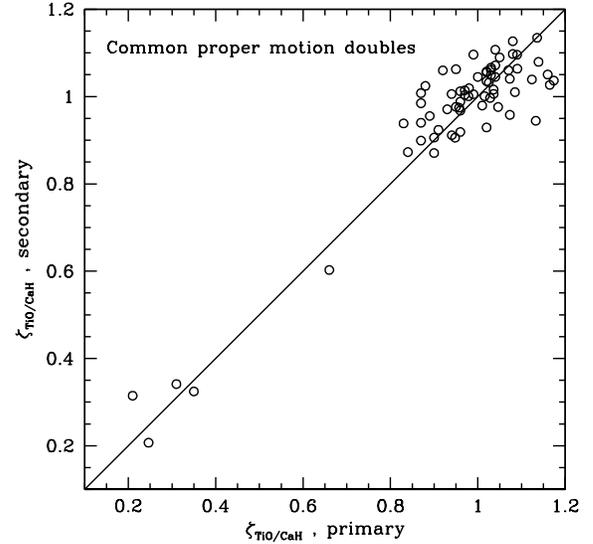}
\caption{Comparison between the $\zeta_{\rm TiO/CaH}$ metallicity
  index in the resolved primaries and secondaries of 71 common proper
  motion doubles. The tight correlation suggests
  that $\zeta_{\rm TiO/CaH}$ is a good measure of relative 
  metallicity in cool stars of various spectral subtypes.}
\end{figure}

We can test the consistency of $\zeta_{\rm TiO/CaH}$ as a metallicity
index by comparing the values of $\zeta_{\rm TiO/CaH}$ measured in
the components of resolved binary systems, where the two stars are
expected to share the same abundance. Our catalog contains 71 pairs of
common proper motion doubles in which both components are in the K5-M9
subtype range. Two of the pairs are subdwarfs, four are extreme
subdwarfs, all others are dwarfs. We compare the values of $\zeta_{\rm
  TiO/CaH}$ for the primaries and secondaries in Figure 5. Values of
$\zeta_{\rm TiO/CaH}$ for the primaries and secondaries are found to
be tightly correlated. This is possible if and only if (1) the
components of high proper motion pairs share the same metallicity, and
(2) the parameter $\zeta_{\rm TiO/CaH}$ is a consistent measure of the
metallicity in low-mass stars of different masses. This suggests a use
for the parameter $\zeta_{\rm TiO/CaH}$ in refining the metallicity
subclass system in low-mass stars, and redefining the subclass
separators in the CaH2+CaH3/TiO5 diagram.
 
\subsection{New metallicity classes: dwarfs, subdwarfs, extreme
  subdwarfs, and ultra subdwarfs}

As demonstrated above, the parameter $\zeta_{\rm TiO/CaH}$ can be used
as a simple yet objective estimate of the metal content in cool stars.
We therefore use it to redefine the traditional metallicity subclasses
(dwarf, subdwarf, extreme subdwarf) of low-mass stars. The fact that
the parameter is based on the widely used CaH2 and CaH3 spectral
indices allows for a direct comparison with earlier work, and the
reclassification of previously observed objects using data already
available in the literature.

Ideally, one would want a new definition which preserves as much
as possible any previous subclass assignment. Unfortunately, the
class separators defined by \citet{BK06} do not strictly run along
lines of constant $\zeta_{\rm TiO/CaH}$ (see below). Hence if we
define a new sdM/esdM subclass separator that mostly agrees with the
old separator at {\em earlier} subtypes (e.g. near
CaH2+CaH3$\simeq1.5$), then most of the stars would fall into the sdM
subclass, with only a few left in the esdM subclass. Such a separation
would not be very useful for the sample at hand. If on the other hand,
we define a new sdM/esdM separator that matches the old one at {\em
  later} subtypes (e.g. near CaH2+CaH3$\simeq0.5$), then the new
separator would put a majority of stars in the esdM class. This would
fail to set up any distinction for the most extreme metal-poor stars
in the sample.

Our solution to this dilemma is (1) to have the new sdM/esdM separator
fall near the old separator at {\em later} subtypes and (2) to
introduce a fourth class of objects which would regroup the most
extreme of the subdwarfs, by the weakness of their TiO
bandstrength. We propose to designate stars from the new, fourth
category as ``ultra subdwarfs''. We find that the introduction of a
fourth subclass gives additional depth to the system and highlights
the subdwarfs with the most extreme properties, which strongly stand
out in the [CaH2+CaH3,TiO5] diagram. We denote stars from the
ultra-subdwarf class as ``usdK'' or ``usdM''.

The new four metallicity subclasses for low-mass stars are formally
defined using our $\zeta_{\rm TiO/CaH}$  metallicity index following:
\begin{eqnarray*}
\zeta_{\rm TiO/CaH}>0.825 \Rightarrow {\rm K-M}\\
0.500<\zeta_{\rm TiO/CaH}<0.825 \Rightarrow {\rm sdK-sdM}\\
0.200<\zeta_{\rm TiO/CaH}<0.500 \Rightarrow {\rm esdK-esdM}\\
\zeta_{\rm TiO/CaH}<0.200 \Rightarrow {\rm usdK-usdM}
\end{eqnarray*}
These separators are drawn in Fig.3 (upper right and lower right
panels). Note that this new formal definition, based on ranges of the
$\zeta_{\rm TiO/CaH}$ parameter, retains some flexibility in that it
allows for possible future refinements in the calibration of $[{\rm
 TiO5}]_{Z_{\odot}}$.

\begin{figure}
\epsscale{2.1}
\plotone{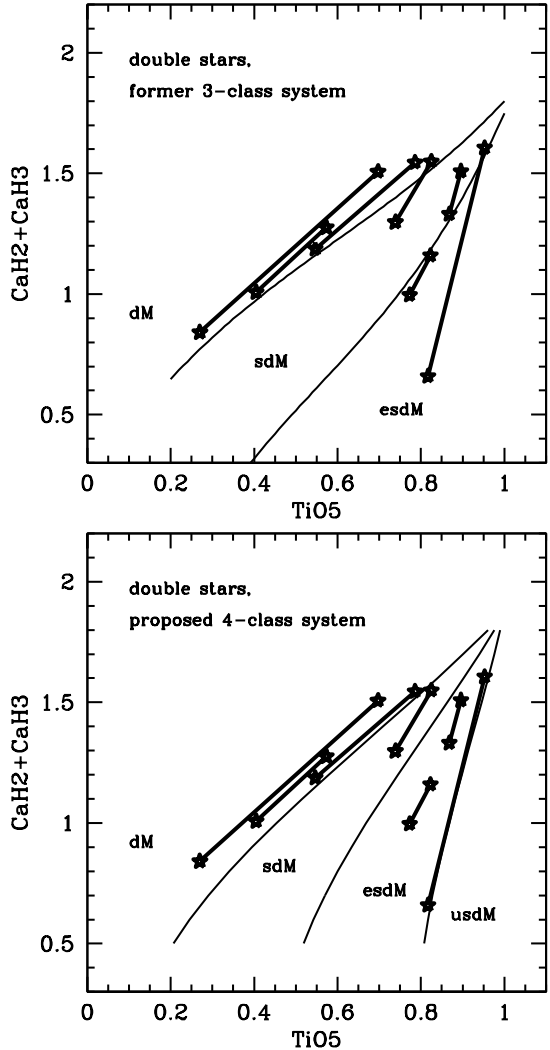}
\caption{Distribution of TiO5 and CaH2+CaH3 index values for members
  of common proper motion pairs. The primaries and secondaries are
  connected by straight lines. In the former classification system, some
  pairs straddle the subclass separators, and have the secondary
  assigned to a different metallicity subclass than the primary. The
  new system more consistently assigns primaries and secondaries to
  the same metallicity class, which is what one should expect assuming
  that components in a pair should have similar abundances.}
\end{figure}

We have already shown in Fig.5 that values of the parameter
$\zeta_{\rm TiO/CaH}$ are tightly correlated for components of common
proper motion pairs. In Figure 6, we use the double stars again to
demonstrate the consistency of the new 4-subclass system. The pairs
are represented in the [CaH2+CaH3,TiO5] diagram. Individual components
are plotted with star symbols, with line segments joining the
components of each pair. The subclass separators from both the old
three-class and new four-class systems are drawn for comparison. 

It is quite apparent from this plot that the new system provides a
more satisfactory separation between the metallicity groups. In the
old system, the segments joining the components clearly do not run
parallel to the subclass separators. One of the pairs straddles the
dM/sdM boundary, with the primary classified as a dwarf and the
secondary and sdM. Another pair has an sdM primary, with a secondary
located deep within the esdM zone. This runs afoul of the assumption
that both components of wide binaries should have similar
abundances. In the new system, however, the separators run nearly
parallel to the double star segments. The most metal-poor of pair
actually has both the primary and secondary sitting on the esdM/usdM
separator, which is precisely what one would expect of the subclass
separators, i.e. that they follow lines of iso-metallicity in the
[CaH2+CaH3,TiO5] plane.

\begin{figure*}
\epsscale{1.0}
\plotone{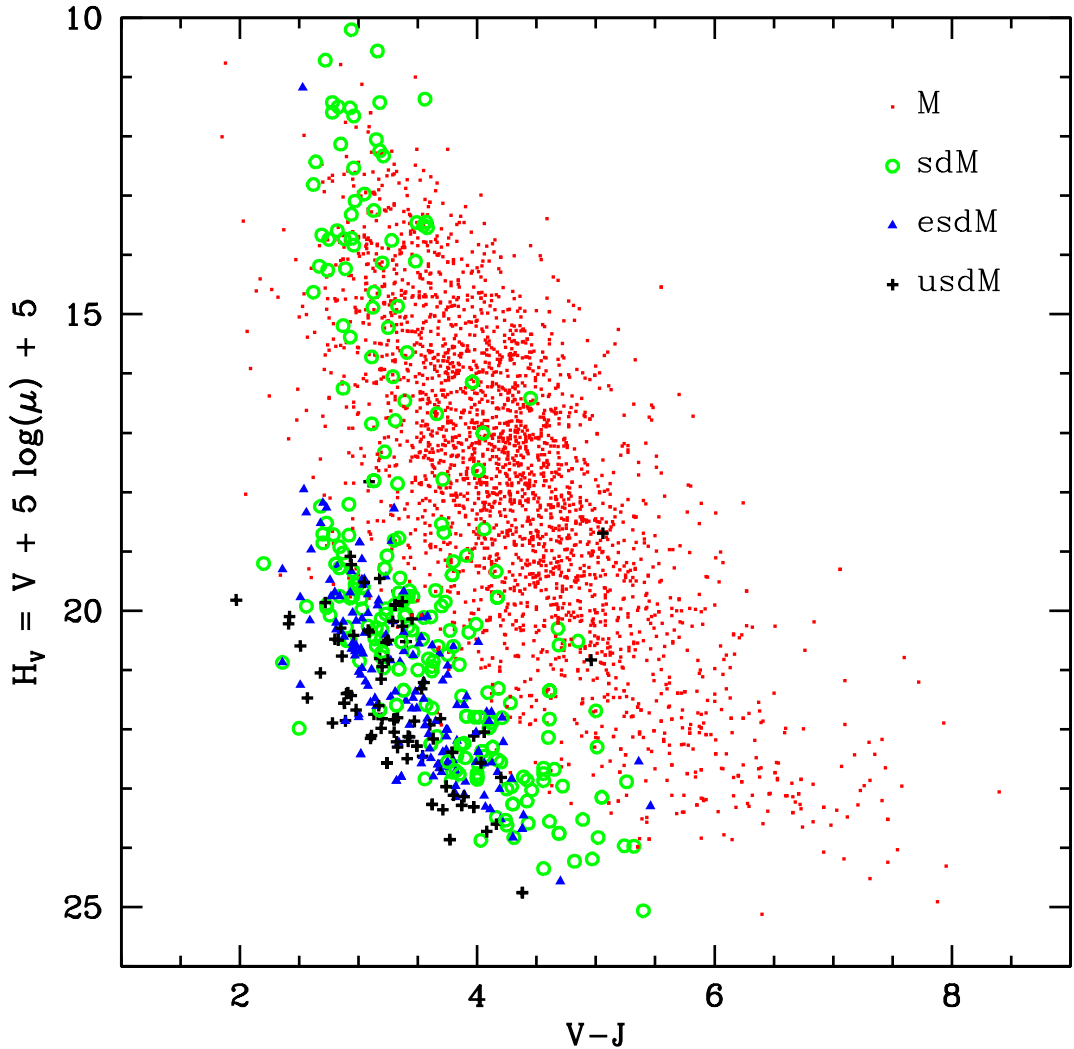}
\caption{Distribution of the spectroscopically confirmed
  dwarfs and subdwarfs in the ($H_V$,V-J) reduced proper motion
  diagram. Dwarfs are plotted as red dots, subdwarfs (sdK,sdM) are
  plotted as open green circles, extreme subdwarfs (esdK,esdM) as
  filled blue triangles, and ultra subdwarfs as black
  crosses. Nearly all subdwarfs fall in the part of the diagram where
  halo stars are expected to cluster (see Fig.2). There is also a
  segregation between the sdM/esdM/usdM stars, with the more
  metal-poor objects having increasingly bluer color and/or larger
  reduced proper motions. A significant number of stars classified as
  early-type M subdwarfs ($V-J<3.5$) are located well within the dwarf
  stars locus, which suggest that at least some of these have been
  misclassified. This is likely due to the limited range of TiO
  bandstrengths in the early-type stars, which makes the metallicity
  subclass assignment more uncertain.}
\end{figure*}

An additional check to the dwarf/subdwarf separator consists of
placing all classified objects back in the reduced proper motion
diagram. We do this in Figure 7, where we plot the stars from each
metallicity subclass in a different symbol/color. One recovers the
striking separation between the dwarfs and subdwarfs, which associates
the dwarfs with the Galactic disk, and the subdwarfs with the Galactic
halo (see Fig.2). The subdwarfs, extreme subdwarfs, and ultra
subdwarfs also show some level of segregation in the reduced proper
motion diagram. The ultra subdwarfs cluster in the extreme lower left
of the diagram, with the extreme subdwarfs and subdwarfs roughly
falling in successive layers above and to the right. This is
consistent with the idea of a dependence of the color-magnitude
relationship on metallicity. It could also indicate that the more
metal-poor stars tend to have larger transverse velocities, and hence
large reduced proper motions. Possibly, both kinematics and
metallicity effects conspire to yield the slightly different observed
loci for the subdwarfs, extreme subdwarfs, and ultra subdwarfs.

One limitation of the classification system become apparent in Fig.7,
however. At the earliest subtypes, i.e. for stars with
CaH2+CaH3$\gtrsim$1.5, the subclass separators converge in the
[CaH2+CaH3,TiO5] plane, and the subclass assignment becomes
increasingly uncertain given the inevitable uncertainty in the
measurement of the spectral indices. The chance of a misclassification
therefore becomes significant for stars of earlier subtypes. As it
turns out, one finds in the reduced proper motion diagram a significant
number of subdwarfs which are well within the normal locus of the M
dwarfs. Because the reduced proper motion is correlated with the
velocity of a star {\em in the plane of the sky}, one would 
expect, statistically, to find a small fraction of halo subdwarfs to have
relatively small reduced proper motions, simply because of a small
{\em projected} velocity. In Fig.7 one actually finds a small
number of late-type ($V-J>3.5$) M subdwarf to hover significantly
above the main halo subdwarf locus, and this can be explained
statistically from projection effects. However, the large clustering
of early-type M subdwarfs located well above the main
subdwarf locus ($H_V<16$) is so clearly distinct that it is unlikely
to be explained by projection effects alone. The suspicion for those
stars is that many of them have probably been misclassified. 

The problem most likely lies in the accuracy of the CaH/TiO bandhead
measurements. With the subclass separators converging in the
[CaH2+CaH3,TiO5] plane for subdwarfs of earlier subtypes, the
metallicity subclass assignment is bound to be increasingly sensitive
on the measurement errors. At this time, so it appears, the accuracy
of our spectral index measurements is not good enough to unambiguously
classify many of the early-type M dwarfs on our list. We therefore
caution that a higher signal-to-noise ratio is required to properly
classify M dwarfs/subdwarfs of earlier spectral subtype.


\section{Comparison with the NextGen model grid}

\begin{figure}
\epsscale{1.1}
\plotone{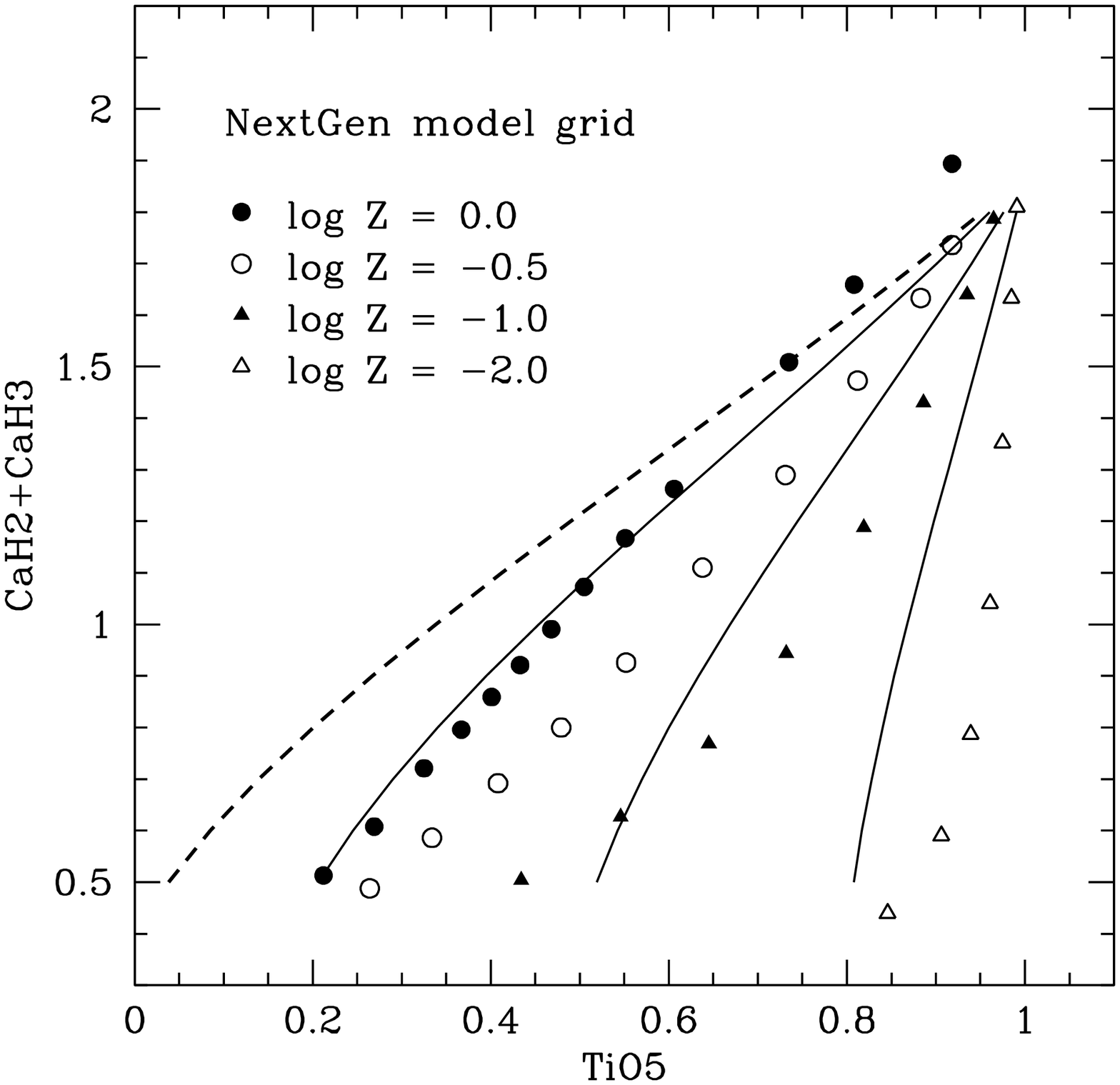}
\caption{Distribution of CaH2+CaH3 and TiO5 indices
  measured for synthetic spectra from the NextGen model grid
  \citep{HAB99}, based on a grid of log g=5.0 models with various
  T$_{eff}$ and metallicities (see legend). Data points are overlaid
  on the the new classification grid, with the dashed line showing the
  observed locus of metal-rich disk stars (log Z=0) and the continuous
  lines mapping the boundaries between the four metallicity
  subclasses, as defined in this paper. The disagreement between the
  NextGen models and our classification scheme is discussed in the text.}
\end{figure}

To estimate the range of [Fe/H] actually associated with each
of the metallicity subclasses, we examine the range of CaH2, CaH3, and
TiO5 spectral indices predicted by the NextGen model atmosphere grid
of \citet{HAB99}. Synthetic spectra for cool stars were retrieved from
the website of the PHOENIX
project\footnote{http://www.hs.uni-hamburg.de/EN/For/ThA/phoenix/}. We
used a grid of spectra with effective temperatures
$3,000<T_eff<4,000$, $log g=5.0$, and metallicities $log
Z=0.0,-0.5,-1.0$, and $-2.0$. The high resolution spectra were
convolved with a Gaussian kernel to match the appearance of the
low-resolution spectra from our spectroscopic follow-up survey. The
distribution of CaH2+CaH3 and TiO5 spectroscopic indices are shown in
Figure 8.

We find that the NextGen grid points overall provide a theoretical
support of only the more general effects of metallicity on the TiO to
CaH bandstrength ratio. The model basically confirms that the lower
the metallicity, the weaker the TiO bandstrength relative to
CaH. However, there are significant discrepancies between the location
of the NextGen grid points and our phenomenological definition of the
metallicity subclasses. The NextGen model grid clearly does not yield
accurate predictions of TiO to CaH ratios. The strongest evidence for
this argument is in the locus of the log Z = 0.0 objects (filled
circles), which should be expected to fall along the empirically
calibrated, mean locus of Galactic disk stars as determined in \S3
(dashed line in Fig.8). The source of this discrepancy is that the TiO
to CaH ratio tends to be overestimated, in the NextGen grid, for the
coolest objects \citep{BCK07}.

Perhaps a cause for more concern is the fact that our subclass
separators do not seem to run parallel to the lines of iso-metallicity
from the NextGen grid. In particular, our subclass separators have
increasing slopes for decreasing values of TiO5, while the NextGen
iso-metallicity grid lines show a decreasing slope for decreasing
values of TiO5. The discrepancy is most pronounced for CaH2+CaH3$<$1. It
is possible that there is an inflection in the locus of the metal-rich
stars due to a saturation of the TiO band, in which case it would be
inappropriate to carry down this inflection to the metal-poor stars
range.

Ultimately, one would want to calibrate the iso-metallicity contours
empirically, perhaps using subgroups of chemically homogeneous,
metal-poor stars, such as groups of cool subdwarfs from globular
clusters. This is unfortunately not possible at this time, for lack of
spectroscopic data on cool subdwarfs in clusters. Metallicity
segregation of field stars based on kinematical selection appears to
be impractical, since all metal-poor stars in our sample at first
glance appear to have similar (halo-like) kinematics and cannot be
separated into distinct kinematical subgroups.

The best method for properly mapping out the lines of iso-metallicity
in the [CaH2+CaH3,TiO5] will probably be to assemble a large sample of
resolved subdwarf binaries, and expand on the procedure illustrated in
Fig.6. A large area of the diagram (0.2$<$TiO5$<$0.7,
0.5$<$CaH2+CaH3$<$0.9) remains non-validated at this time. Efforts
should be devoted to identifying additional pairs of common proper
motion subdwarfs, particularly pairs for which at least one component
has CaH2+CaH3$<$1.0, which would correspond to a spectral subtype
later than $\approx$sdM4.0/esdM4.0/usdM4.0. Accurate measurement of
spectral indices in such double stars would be critical in refining
the cool subdwarf classification system.

\section{Determination of spectral subtypes in the cool subdwarfs}

The assignment of spectral subtypes for cool subdwarfs and extreme
subdwarfs is traditionally based on the depth of the CaH molecular
bands. The depth of TiO bands are widely used in the classification of
low-mass dwarfs, but TiO bandstrengths show too much dependence on the
metallicity to be useful in the subdwarf regime. In the classification
system of G97, spectral subtypes of subdwarfs and extreme subdwarfs are
determined from the CaH2 and CaH3 indices using these relationships:
\begin{eqnarray*}
Sp_{\rm CaH2}=  7.91 \ {\rm CaH2}^2 - 20.63 \ {\rm CaH2} + 10.71 \\
Sp_{\rm CaH3}|_{sd} =  13.78 - 16.02 \ {\rm CaH3} \\
Sp_{\rm CaH3}|_{esd}=  11.50 - 13.47 \ {\rm CaH3}
\end{eqnarray*}
Spectral subtypes for the subdwarfs ($Sp_{sd}$) and for the extreme
subdwarfs ($Sp_{esd}$) are thus formally assigned using slightly
different equations:
\begin{eqnarray*}
Sp_{sd} = 0.5 ( Sp_{\rm CaH2} + Sp_{\rm CaH3}|_{sd} ) \\
Sp_{esd} = 0.5 ( Sp_{\rm CaH2} + Sp_{\rm CaH3}|_{esd} ) .
\end{eqnarray*}
The reason why separate relationships were used for dwarfs and
subdwarfs has to do with the fact that the CaH2 and CaH3 indices are
not strictly correlated to each other, and in fact their ratio appears
to be a function of metallicity. The $Sp_{\rm CaH2}$ relationship was
actually borrowed from the spectral subtype relationship of dwarf
stars and, as a result, distinct $Sp_{\rm CaH3}$ relationships had to
be defined for the subdwarfs and extreme subdwarfs.

\begin{figure*}
\epsscale{1.15}
\plotone{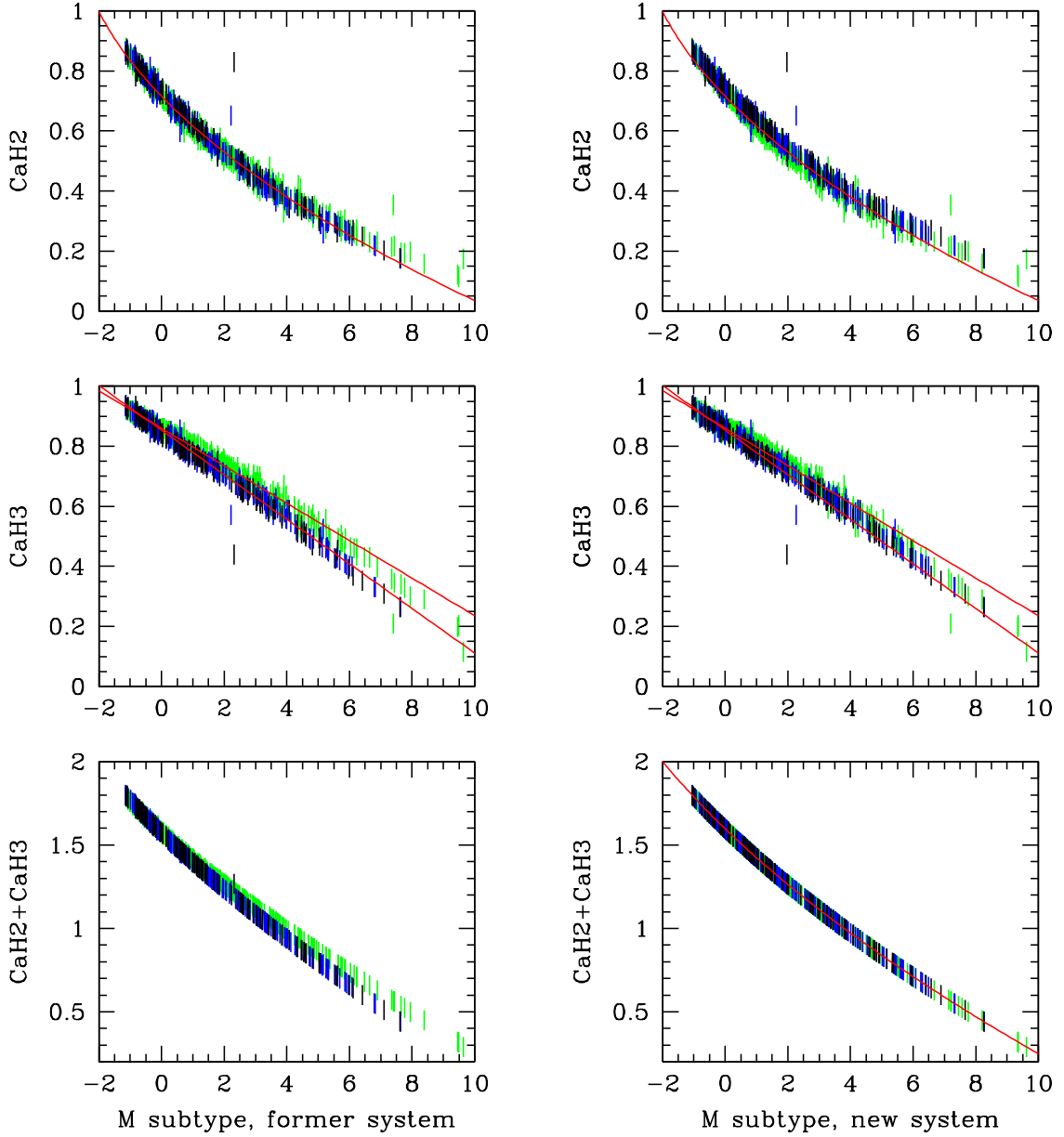}
\caption{Left panels: variation in the CaH spectral indices with spectral 
subtype, for subtypes assigned in the system of G97. Separate 
plots are shown for the CaH2 index (top), the CaH3 index (middle), and the 
combined CaH2+CaH3 index (bottom). For more clarity, subtype 
values have {\em not} been rounded up to the nearest half subtype, as 
would be customary. Subdwarfs are shown in green, extreme subdwarfs in
blue, ultra subdwarfs in black. The G97 system assigns
subtypes based on the average of a set of relationships using the CaH2
and CaH3 indices, and  uses different relationships for subdwarfs and
extreme subdwarfs (red curves); note the slight discrepancy for
subdwarfs of later subtypes. Right panels: variation in the CaH
spectral indices with spectral subtype, for subtypes assigned
according to Equation 4 in this paper. Separate plots are shown for
the CaH2 index (top), the CaH3 index (middle) and the combined
CaH2+CaH3 index, on  which Eq.5 is based (bottom). The new
relationship for subtype assignment (red curve), simplifies the
classification system as the same relationship is used for all
subclasses. The most significant results is in the subtype assignment
of late esdM/usdM, which are classified 0.5-1.0 subtype later in the
new system.}
\end{figure*}

The relationships defined in G97 were based on the original
spectroscopic sample of only 79 objects, a sample much
smaller than the one currently available. This original sample was
particularly deficient in late-type stars, with only two sdM5.0 and
one sdM7.0 as the latest subdwarfs, and an esdM5.5 as the latest
extreme subdwarf. Cooler object which were discovered afterward, were
classified from a simple extrapolation of the relationships
above. Figure 9 shows the variation of the CaH2, CaH3, and combined
CaH2+CaH3 indices as a function of assigned spectral subtypes. Red
continuous lines show the assumed relationship between the indices and
the spectral subtypes, in the G97 system.

The assignment of spectral subtypes in the new four-class system
introduces a conundrum. Spectral subtyping for the dwarf stars
is already very well established, and needs not be changed. But the
spectral subtypes for three subdwarf classes is evidently in need of
revision, because the original sdM and esdM classes used slightly
different relationships and because the new sdM/esdM/usdM subclasses
do not have a strict correspondence to the older ones. One solution is
to retain the use of $Sp_{\rm CaH2}$ as a fiducial, and redefine new
$Sp_{\rm CaH3}$ calibrations for each of the sdM, esdM, and usdM
subclasses. However, the reason why CaH2 and CaH3 are not strictly
correlated comes from the fact that the CaH2 index is defined in a
region of a CaH band which is also overlapped with a major TiO band
extending redward from $\approx6700$\AA. This is most obvious in the
metallicity sequence show in Fig.4, and implies that values of the
CaH2 index must be a function of metallicity in the subdwarfs. Hence
using a metallicity-dependent index to define the spectral subtypes
does not sound like the best choice.

\begin{figure}
\epsscale{1.1}
\plotone{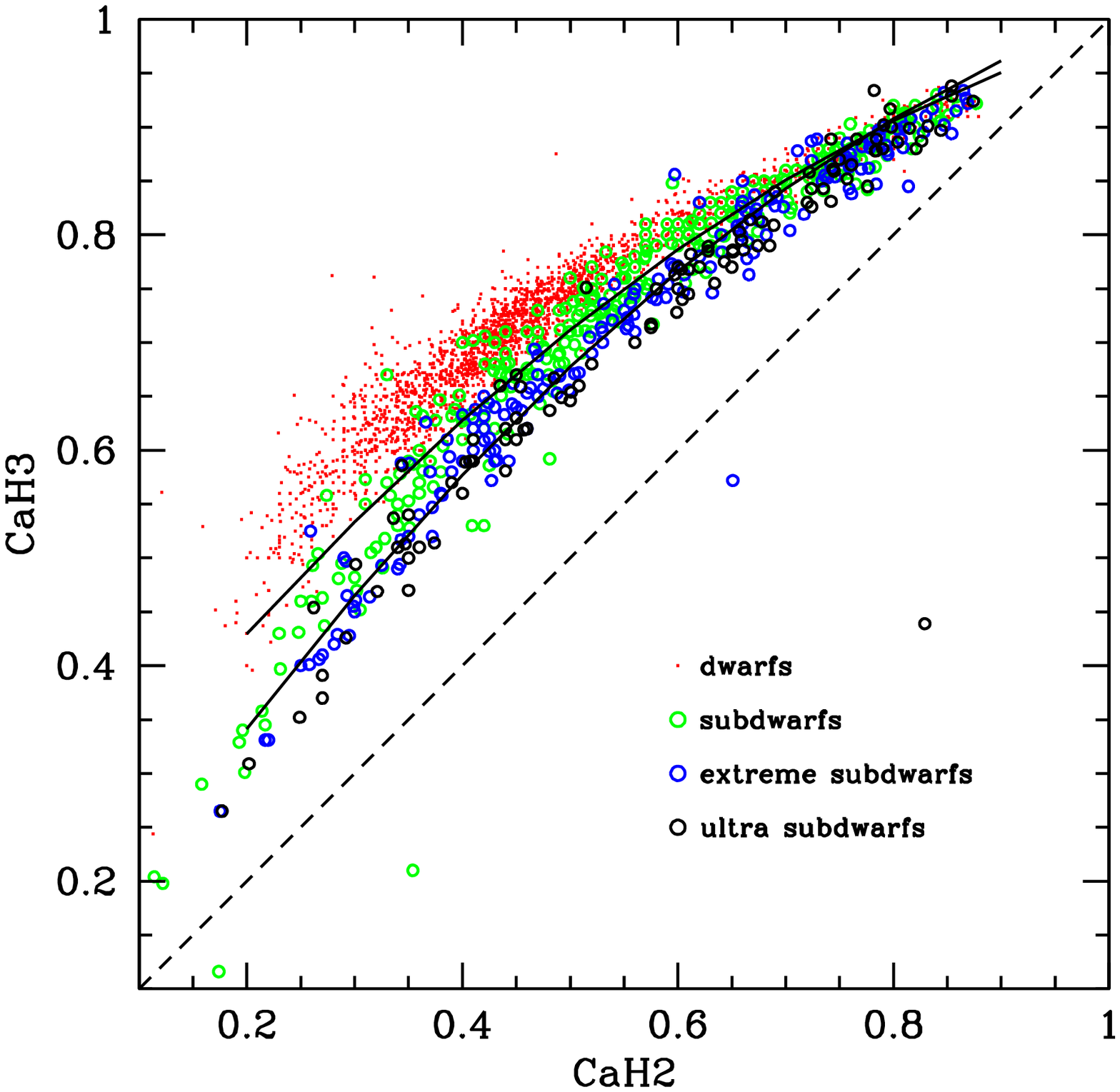}
\caption{Distribution of the CaH2 and CaH3 spectral indices for dwarfs
  and subdwarfs. Objects are segregated according to their metallicity, 
  with the dwarfs in particular occupying a locus up and to the left of 
  the distribution. The distribution suggests that at least one of the two
  indices is dependent of metallicity. The likely culprit is CaH2,
  which probably includes a contribution from a nearby TiO band, and 
  thus presents a deeper band in more metal rich stars.}
\end{figure}

Another solution might be to redefine the spectral subtypes based on
the strength of the CaH3 index alone, which is not so obviously
dependent on the metallicity. Unfortunately, the CaH3 index doesn't
constrain the subtype sequence as well as CaH2 for earlier spectral
subtypes. This is clear from Figure 10, which shows the observed
relationship between the CaH2 and CaH3 indices. The CaH2 index
decreases faster than CaH3 for index values larger than
$\approx0.5$. The trend is reversed for late-type stars, because the
CaH2 gets close to saturation and slowly becomes degenerate. While it
would be tempting to use the CaH2 index in early-type objects and the
CaH3 index in late-type objects, this would make spectral subtype
assignment non-straightforward, as it would require pre-assigning a
spectral subtype.

Our best compromise is to assign spectral subtypes simply based on the
combined value of the CaH2 and CaH3 indices. The use of CaH2+CaH3 also
fits well with our metallicity subclass assignment, which is also
dependent on CaH2+CaH3. We find that we can use the relationship:
\begin{equation}
Sp = 1.4 ({\rm CaH2+CaH3} )^2 - 10.0 ({\rm CaH2+CaH3} ) + 12.4,
\end{equation}
which yields a spectral subtype assignment which very closely fits
the assignment in the G97 system, with only minor differences.
We adopt Equation 4 as a general relationship for assigning spectral
subtypes for subdwarfs, extreme subdwarfs, and ultra subdwarfs. Panels
on the right in Fig.9 show how CaH2 (top), CaH3 (middle), and
CaH2+CaH3 (bottom) vary within the new spectral subtype system.

The main difference between the old and new systems occurs at the
latest subtypes for extreme subdwarfs, which tend to be classified
with 0.5-1.0 subtypes later in the new system. Hence the star 2MASS
1227-0447, classified as esdM7.5 by \citet{BCK07}, is now classified
as usdM8.5 in our new system, using the same spectral index values.

\section{Classification standards for the subdwarf, extreme subdwarf,
  and ultra subdwarf sequences.}

We are now assigning metallicity subclasses and spectral subtypes to
all the stars observed in our spectroscopic survey using the new
four-class system described above. In its current form, the new
classification system remains entirely based on values taken by the
three spectroscopic indices CaH2, CaH3, and TiO5, as in the former
3-class system of G97.

Traditionally, however, formal spectral classification has
often been determined by comparing a target to a set of reference
objects formally defined as ``classification standards". The use of
classification standards makes possible the determination of spectral
subtypes using any other part of the spectral energy distribution, by
direct comparison with the standards, thus possibly freeing one from
the need to always obtain spectra of the same narrow spectral
region. This is the essence of the MK process, as elaborated by, e.g.,
\citet{MK73}.

From our census of high proper motion subdwarfs, we have assembled
sequences of objects to be used as standard sequences for the
classification of subdwarfs, extreme subdwarfs, and ultra
subdwarfs. These standards were selected based on the following
criteria: (1) a spectrum representative of their specific subclass and
subtype, (2) a star among the brightest objects of their respective
subclass and subtype, (3) an available spectrum with the best possible
signal-to-noise ratio, and (4) a location of the sky
$-30^{\circ}<$Decl.$<+30^{\circ}$ to make the star visible from both
hemispheres. We tried to follow those guidelines as much as possible,
only skipping stars from the literature for which we could not obtain
a copy of the original spectra. A list of the proposed standards is
given in Table 2.

\begin{figure*}
\epsscale{1.0}
\plotone{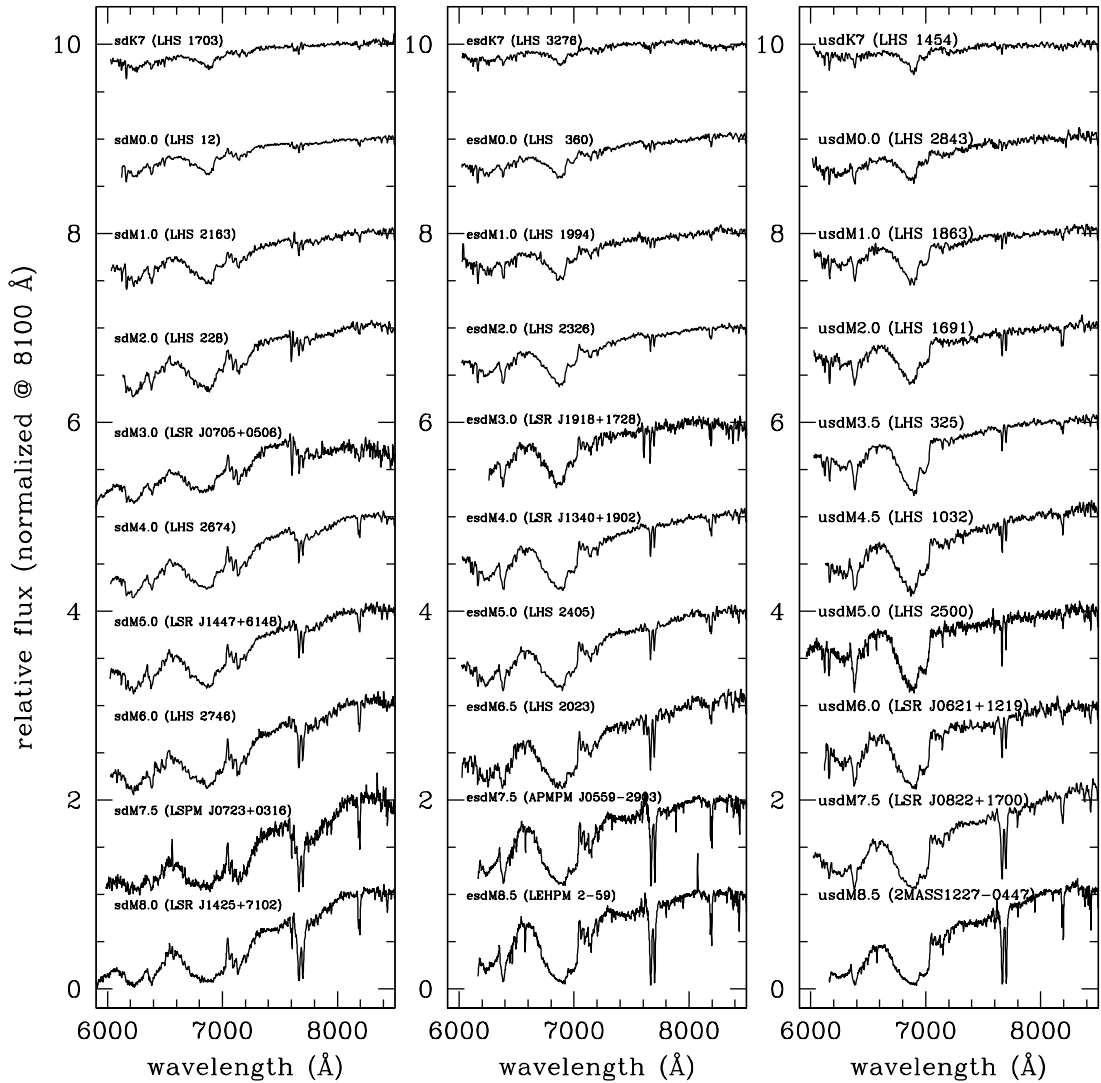}
\caption{Sequences of classification standards for low-mass subdwarfs
  (left), extreme subdwarfs (center), and ultra subdwarfs
  (right). Data on those stars are found in Table 2. Metallicity
  classes (sd, esd, usd) are based on the ratio of the TiO bandhead
  blueward of 7000\AA to the CaH bandhead redward of 7000\AA (see
  \S4). All the spectra shown are from our follow-up survey, except
  spectra from the esdM7.5 star APMPM J0559-2903, the esdM8.5 star
  LEHPM 2-59, and the usdM8.5 star 2MASS1227-0447, all three
  graciously provided by A. Burgasser.}
\end{figure*}

Spectra of all standards are displayed in Figure 11. Our standard
sequence for subdwarfs (sdM) spans the range from sdK7 to sdM8.0. The
sequences for the extreme subdwarfs (esdM) extends to esdM8.5, and the
ultra subdwarf (usdM) sequence extends to usdM8.5. A handful of stars
of even later subtypes  have been reported in the literature
\citep{SLM04,BCK07} which could be used to extend the sequence into
the L subdwarfs regime. Because of the paucity of such extremely cool
stars, we refrain from extending the sequence until more data has been
assembled. In any case, it is not obvious that the relationships
defined here for the classification of earlier objects should apply to
the coolest M dwarfs, and into the L subdwarfs regime. In the
ultra-cool dwarfs and subdwarfs, condensate formation depletes Ca and
Ti, and the deepening and broadening of the K I absorption doublet
obscures the 7000\AA region \citep{Ketal99}. Other features in the red
optical spectra would probably be more useful in extending the scheme
to later objects (e.g. Kirkpatrick et al. 1995).

It should also be noted that at the latest subtypes, the CaH index is
much more sensitive to variations in metallicity. It has been
suggested that for the ultra-cool subdwarfs, a {\em mild   subdwarf}
class (denoted d/sd) could be defined \citep{BCK07}; the high proper
motion star SSSPM 1444-2019, tentatively classified as d/sdM9, would
be the prototype for this intermediate class of objects. The colors of
late subdwarfs show a wide distribution which appear to be a function
of metallicity \citep{SLM04b}; color-color scheme might prove useful
in further refining or ``fine-tuning'' the metallicity subclasses for
cooler objects.

\section{Conclusions}

We have redefined the three metallicity subclasses of low-mass
stars of spectral type M based on a recalibration of the TiO to CaH
bandstrength ratio, in the spectral region around 7000\AA. The new
definition retains the dwarf (K5-M9), subdwarf (sdK5-sdM9) and extreme
subdwarf (esdK5-esdM9) classes, but introduces a new subclass of
objects, the ultra subdwarfs (usdK5-usdM9), which comprise all the
stars with the smallest observed apparent metallicity. The different
subclasses are redefined based on the ratio of the TiO to CaH
bandstrengths, which is correlated with metallicity. The TiO to CaH
ratio is calibrated using a subsample of kinematically selected disk
stars of presumably solar abundances. A metallicity parameter,
$\zeta_{\rm   TiO/CaH}$ (see Eq.4), provides a numerical estimate of
how the TiO to CaH ratio in a star compares to the value measured in
solar-metallicity objects. Stars with close to solar abundances have
$\zeta_{\rm TiO/CaH}\simeq1$, by definition, while metal-poor stars
have $\zeta_{\rm TiO/CaH}<1$, with a value close to 0 representing
stars  with virtually no detectable TiO molecular absorption,
indicative of an ultra-low metal content.

We have also redefined the assignment of the spectral subtypes in all
three metallicity sequences based on the sum of the CaH2 and CaH3
indices (CaH2+CaH3). We introduce three sequences of representative
objects for each of the three metallicity classes, covering the range
of spectral subtypes. These objects should be useful as spectroscopic
classification standards. Most standards are located in the northern
sky, between declinations 0 and +30. Future efforts should be invested
in finding additional standard objects at southern declinations.

Ultimately, the use of the TiO/CaH bandstrength ratio as an effective
measure of the metallicity in low-mass stars, through the parameter
$\zeta_{\rm TiO/CaH}$ should be vetted against independent
measurements of the metallicity in selected low-mass, metal-poor
stars. Such critical metallicity calibration has now been initiated
based on atmospheric modeling of Fe lines using high resolution
spectra of bright (mostly early-type) M dwarfs and subdwarfs
\citep{WW05,WW06,BSHGB06}. A calibration of the $\zeta_{\rm TiO/CaH}$
parameter with metallicity would be most useful as it would provide a
simple means to measure the metallicity in low-mass stars using
spectra of only moderate spectral resolution. Abundances could thus
potentially be determined for large numbers of distant, faint,
low-mass subdwarfs from the Galactic halo, which would become
formidable probes of ancient Galactic formation and evolution.


\acknowledgments

{\bf Acknowledgments}

The authors would like to thank Adam Burgasser for a critical reading
of the manuscript. His comments and suggestions resulted in
significant improvements of the paper.

This research program was supported by NSF grant AST-0607757 at
the American Museum of Natural History. SL and MS also gratefully
acknowledge support from Mr. Hilary Lipsitz.

\clearpage
\begin{landscape}
\begin{deluxetable}{cllccrrcccccrr}
\tabletypesize{\scriptsize}
\tablewidth{0pc} 
\tablecolumns{14} 
\tablecaption{Classification standards for subdwarfs (sd), extreme
  subdwarfs (esd), and ultra-subdwarf (usd).}
\tablehead{
\colhead{Spectral} & 
\colhead{LSPM} &
\colhead{other} &
\colhead{R.A.} &
\colhead{Decl.} &
\colhead{$\mu$R.A.\tablenotemark{a}} &
\colhead{$\mu$Decl.\tablenotemark{a}} &
\colhead{V\tablenotemark{b}} &
\colhead{V-J\tablenotemark{c}} &
\colhead{CaH2\tablenotemark{d}} &
\colhead{CaH3\tablenotemark{d}} &
\colhead{TiO5\tablenotemark{d}} &
\colhead{$\zeta_{\rm TiO/CaH}$} &
\colhead{ref.\tablenotemark{e}} \\
\colhead{type} &
\colhead{catalog \#} &
\colhead{name} &
\colhead{J2000} &
\colhead{J2000} &
\colhead{$\arcsec$ yr$^{-1}$} &
\colhead{$\arcsec$ yr$^{-1}$} &
\colhead{mag} &
\colhead{mag} &
\colhead{} &
\colhead{} &
\colhead{} &
\colhead{} &
\colhead{}
}
\startdata 
\cutinhead{Subdwarfs}
 sdK7.0 & LSPM J0448+2206 &       LHS 1703 & 04 48 32.21 & +22 06 25.0 & 0.047&-0.696& 14.51 & 2.92 & 0.851 & 0.911 & 0.952 & 0.64 & L07 \\
 sdM0.0 & LSPM J0202+0542 &       LHS   12 & 02 02 52.15 & +05 42 20.5 & 2.325&-0.714& 12.28 & 2.81 & 0.749 & 0.881 & 0.876 & 0.60 & L07 \\
 sdM1.0 & LSPM J0828+1709 &       LHS 2163 & 09 38 17.53 & +22 00 43.6 &-0.742& 0.015& 14.33 & 3.72 & 0.631 & 0.811 & 0.740 & 0.76 & L07 \\
 sdM2.0 & LSPM J0716+2342 &       LHS  228 & 07 16 27.71 & +23 42 10.4 & 0.940&-0.584& 15.48 & 3.46 & 0.520 & 0.742 & 0.714 & 0.62 & L07 \\
 sdM3.0 & LSPM J0705+0506 & LSR J0705+0506 & 07 05 48.76 & +05 06 17.1 & 0.185&-0.473& 16.42 & 2.73 & 0.421 & 0.680 & 0.592 & 0.70 & L03b \\
 sdM4.0 & LSPM J1303+2328 &       LHS 2674 & 13 03 34.84 & +23 28 46.3 &-0.452&-0.415& 16.36 & 4.68 & 0.382 & 0.604 & 0.554 & 0.65 & L07 \\
 sdM5.0 & LSPM J0852+1530 & LSR J1447+6148 & 14 48 46.87 & +61 48 02.6 & 0.150&-0.966& 19.40 & 4.56 & 0.315 & 0.505 & 0.452 & 0.69 & L03b \\
 sdM6.0 & LSPM J1331+2447 &       LHS 2746 & 13 31 28.24 & +24 47 10.9 &-0.497&-0.225& 20.07 & 4.69 & 0.305 & 0.452 & 0.383 & 0.75 & L07 \\
 sdM7.5 & LSPM J0723+0316 &        \nodata & 07 23 43.06 & +03 16 21.8 &-0.102&-0.401& 19.80 & 5.26 & 0.193 & 0.329 & 0.271 & 0.77 & L07 \\
 sdM8.0 & LSPM J1425+7102 & LSR J1425+7102 & 14 25 05.03 & +71 02 09.6 &-0.618&-0.170& 19.79 & 5.02 & 0.198 & 0.301 & 0.332 & 0.73 & L03a \\
\cutinhead{Extreme subdwarfs}					      						
esdK7.0 & LSPM J1715+3037 &       LHS 3276 & 17 15 46.65 & +30 37 57.5 &-0.175&-0.706& 14.66 & 2.60 & 0.854 & 0.894 & 0.967 & 0.39 & L07 \\
esdM0.0 & LSPM J1346+0542 &       LHS  360 & 13 46 55.52 & +05 42 56.4 &-0.766&-0.851& 15.43 & 3.04 & 0.734 & 0.850 & 0.929 & 0.34 & L07 \\
esdM1.0 & LSPM J0814+1501 &       LHS 1994 & 08 14 33.54 & +15 01 07.8 & 0.293&-0.409& 16.63 & 3.03 & 0.665 & 0.793 & 0.883 & 0.35 & L07 \\
esdM2.0 & LSPM J1054+2406 &       LHS 2326 & 10 54 25.89 & +24 06 44.5 &-0.455&-0.490& 16.46 & 3.62 & 0.553 & 0.723 & 0.847 & 0.32 & L07 \\
esdM3.0 & LSPM J1918+1728 & LSR J1918+1728 & 19 18 36.99 & +17 28 00.2 &-0.131&-0.621& 19.05 & 3.81 & 0.447 & 0.662 & 0.837 & 0.28 & L03b \\
esdM4.0 & LSPM J1340+1902 & LSR J1340+1902 & 13 40 40.68 & +19 02 23.1 &-0.455&-0.803& 18.47 & 5.46 & 0.381 & 0.558 & 0.799 & 0.28 & L03b \\
esdM5.0 & LSPM J0843+0600 &       LHS 2405 & 08 43 58.45 & +06 00 39.4 & 0.314&-0.393& 18.50 & 3.74 & 0.342 & 0.494 & 0.758 & 0.31 & L07 \\
esdM6.5 & LSPM J0830+3612 &       LHS 2023 & 08 30 51.65 & +36 12 57.0 & 0.398&-0.701& 18.27 & 3.36 & 0.267 & 0.406 & 0.708 & 0.34 & L07 \\
esdM7.5 &         \nodata &APMPM J0559-2903& 05 58 58.64 & -29 03 27.2 & 0.370& 0.060& 19.50 & 4.61 & 0.217 & 0.331 & 0.604 & 0.42 & B06 \\
esdM8.5 &         \nodata &      LEHPM 2-59& 04 52 09.94 & -22 45 08.4 & 0.064& 0.746& 19.10 & 3.60 & 0.175 & 0.265 & 0.656 & 0.35 & B06 \\
\cutinhead{Ultra subdwarfs}					      						
usdK7.0 & LSPM J0251+2442 &       LHS 1454 & 02 51 14.69 & +24 42 47.1 & 0.558&-0.296& 16.22 & 2.41 & 0.844 & 0.897 & 0.998 & 0.02 & L07 \\
usdM0.0 & LSPM J1401+0738 &       LHS 2843 & 14 01 13.75 & +07 38 08.8 &-0.528&-0.696& 16.68 & 2.91 & 0.742 & 0.889 & 1.003 &-0.02 & L07 \\
usdM1.0 & LSPM J0642+6936 &       LHS 1863 & 06 42 05.77 & +69 36 46.7 &-0.359&-0.382& 16.76 & 3.09 & 0.651 & 0.786 & 1.002 &-0.01 & L07 \\
usdM2.0 & LSPM J0440+1538 &       LHS 1691 & 04 40 14.52 & +15 38 51.9 & 0.087&-0.581& 18.27 & 3.97 & 0.575 & 0.717 & 1.019 &-0.04 & L07 \\
usdM3.5 & LSPM J1221+2854 &       LHS  325 & 12 21 32.83 & +28 54 22.7 &-1.019&-0.295& 16.99 & 3.42 & 0.440 & 0.581 & 0.994 & 0.01 & L07 \\
usdM4.5 & LSPM J0011+0420 &       LHS 1032 & 00 11 00.78 & +04 20 25.0 & 0.152&-0.515& 18.40 & 4.06 & 0.374 & 0.514 & 0.944 & 0.08 & L07 \\
usdM5.0 & LSPM J1202+1645 &       LHS 2500 & 12 02 32.29 & +16 45 35.3 &-0.818&-0.188& 19.24 & 3.77 & 0.321 & 0.469 & 0.978 & 0.03 & L07 \\
usdM6.0 & LSPM J0621+1219 & LSR J0621+1219 & 06 21 34.48 & +12 19 43.7 & 0.268&-0.441& 18.36 & 3.60 & 0.292 & 0.426 & 0.971 & 0.03 & L03b\\
usdM7.5 & LSPM J0822+1700 & LSR J0822+1700 & 08 22 33.75 & +17 00 18.9 & 0.359&-0.494& 19.80 & 4.08 & 0.202 & 0.309 & 0.887 & 0.12 & L04 \\
usdM8.5 &         \nodata & 2MASS1227-0447 & 12 27 05.06 & -04 47 20.7 &-0.459& 0.249& 19.40 & 3.91 & 0.177 & 0.265 & 0.825 & 0.18 & B07 \\
\enddata
\tablenotetext{a}{Positions at the 2000.0 epoch and in the J2000
  equinox.}
\tablenotetext{b}{Optical V magnitudes estimated from combined
  photographic blue and red magnitudes in the Digitized Sky Survey,
  see \citet{LS05}.}
\tablenotetext{c}{Infrared J magnitudes from the 2MASS all-sky catalog
  of point sources \citep{C03}.}
\tablenotetext{d}{Molecular band indices as defined in \citet{RHG95}.}
\tablenotetext{e}{Source of the spectroscopy: L03a \---
  \citet{LSR03a}, L03b \--- \citet{LRS03b}, L04 \--- \citet{LSR04},
  B06 \--- \citet{BK06}, B07 \--- \citet{BCK07},  L07 \--- Lepine et
  al. (2007) in preparation.}
\end{deluxetable} 
\clearpage
\end{landscape}

\end{document}